\documentclass{aa}  

\bibpunct{(}{)}{;}{a}{}{,} 
\usepackage{graphicx}
\usepackage{txfonts}
\usepackage{hyperref}
\def\der{{\rm d}}
\begin{document} 
\renewcommand{\arraystretch}{1.5}
   \title{Constraining Redshifts of Unlocalised Fast Radio Bursts}

   \author{Charles R. H. Walker,
          \inst{1}
          Yin-Zhe Ma,
          \inst{2,3,1}
          \and
          Rene P. Breton\inst{1}
          }

   \institute{Jodrell Bank Center for Astrophysics, The University of Manchester,
              Alan Turing Building, Oxford Road, Manchester, M13 9PL, UK\\
              \email{charles.walker@postgrad.manchester.ac.uk, rene.breton@manchester.ac.uk}
         \and
         School of Chemistry and Physics, University of KwaZulu-Natal, Westville Campus, Private Bag X54001, Durban, 4000, South Africa 
           \\
             \email{ma@ukzn.ac.za}
          \and   
            NAOC-UKZN Computational Astrophysics Center (NUCAC),
             University of KwaZulu-Natal, Durban, 4000, South Africa
             }

   \date{\today}

 
  \abstract
   {The population of fast radio bursts (FRBs) will continue to diverge into two groups depending on their method of discovery: those which can be localised, and those which cannot. New projects such as ASKAP and MeerKAT present excellent opportunities to localise many FRBs. However, events potentially less useful for astronomical and cosmological purposes due to limited localisation will continue to accumulate with the advent of new facilities and continued efforts by, for example, the SUPERB collaboration, both of which may require afterglows or multi-wavelength counterparts for sub-arcsecond localisation. It is important to exploit these sources to their maximum scientific potential.}
   {We perform comprehensive analysis of FRB dispersion measures (DMs), considering different theoretical FRB progenitors with view to place more rigorous constraints on FRB redshifts, in particular for large statistical samples, via their DMs.}
   {We review the components which make up FRB DMs, and build redshift-scalable probability distributions corresponding to different progenitor scenarios. We combine these components into a framework for obtaining FRB DM probabilities given their redshifts. Taking into account different possibilities for the evolution of progenitors across cosmic time we invert this model, allowing derivation of redshift constraints.}
   {Effects of varying FRB progenitor models are illustrated. While, as expected, host galaxy DM contributions become decreasingly important with increasing redshift, for AGN-like progenitor scenarios they could remain significant out to redshift 3. Constraints are placed on redshifts of all catalogued FRBs to date with various models and increasingly realistic models may be employed as general understanding of FRBs improves. For localised FRBs, we highlight future prospects for disentangling host and intergalactic medium DM components using their respective redshift scaling. We identify a use for large samples of unlocalised FRBs resulting from upcoming flux-limited surveys, such as with CHIME, in mapping out the Milky Way contribution to the DM.}
   {}


   \maketitle
%

\section{Introduction}

First discovered by \citet{lori07a}, fast radio bursts (hereafter FRBs) are short-duration (approximate width 0.1-10 ms) radio bursts of extragalactic origin. Currently, tens of catalogued FRBs exist in literature \citep{petr16a}\footnote{\url{http://www.frbcat.org}}. FRBs are almost exclusively observed as isolated events: save for a single source detected \citep{spit14a} and then seen to repeat \citep{spit16a}, the observed population consists of lone, individual pulses distributed randomly across the sky.

The extragalactic origin of FRBs may be confirmed or inferred by one of two methods: using interferometry, or via dispersion measure. In the case of a repeater such as FRB121102 where a host galaxy may be unambiguously identified, a spectroscopic redshift for the host, and thus the FRB, may be obtained \citep{tend17a}. The majority of discovered FRBs, however, remain unconstrained to within the large fields of view of their telescopes of detection. The Parkes 13-beam receiver has yielded the majority of FRB discoveries to date but at L-band (1.4 GHz) a single beam has a field of view $\sim$0.55 square degrees \citep{stav96a}, which could contain up to $\sim$83\,0000 galaxies (to a limiting magnitude of I$_{AB}\,(10\sigma)<27.5$) \citep{scov07a}. A burst detected in such a field of view may not be unambiguously associated with a single source. Distances to such bursts are inferred via the measurement of the integrated column density of electrons along the line of sight, termed dispersion measure (hereafter DM). The extragalactic nature of such FRBs are implied via DMs in excess of those possible through the contribution of Milky Way (hereafter MW) electrons alone \citep{lori07a}.

Standard analysis of FRB distances involves subtraction of ${\rm DM}_{\rm MW}$, the DM portion attributed to electrons in the MW along the line of sight, from the total observed DM, leaving an excess, ${\rm DM}_{\rm exc}$ from which a redshift is estimated \citep{lori07a}. This method assumes no DM contribution from the FRB host galaxy or burst environment and yields only upper redshift limits \citep{petr16a}. During post-discovery analysis, authors have previously considered host galaxy DM contributions of $\sim$100 pc\;cm$^{-3}$ \citep[see, e.g.,][]{lori07a, thor13a}, resulting in alternative lower redshift estimates, however cosmological time dilation suppresses high-redshift host DM contributions \citep{zhou14a}, and often no uncertainties are considered alongside either method.

Large quantities of FRBs with identified redshifts could have significant uses in astronomy \citep{macq15a}: binned as a function of redshift, 10$^4$ FRBs could be used to trace the Universe's missing baryonic matter; 10$^{3}$ FRBs could also be used as a cosmic ruler to determine the geometry of the Universe out past redshift 2. However, it has been noted that key to this potential is accurate knowledge of FRB redshifts, which requires details of host galaxy contributions to their total DMs (see, e.g., \citet{macq15a} for a review).

So far, one FRB has been irrefutably localized. As observatories develop the capability to spot in real time and localize sources, the number of FRBs suitable for cosmological purposes will increase. The Canadian Hydrogen Intensity Mapping Experiment (CHIME), for example, will improve localization for potentially tens of FRBs a day within its $\sim250$ square degree field of view by synthesising up to 1024 beams \citep{ng17a}. The large fields of view and on-sky time available to resolution-limited single-dish surveys suggests that many FRBs may still be discovered to worse than sub-arcsecond accuracy and may never have precise localization. To take full advantage of the total FRB population and enable statistical analysis without accurate redshifts it will become increasingly important to understand these sources' DM contributions.

The observed DM of FRBs consists of a combination of components: from the MW, the intergalactic medium (hereafter IGM), the FRB host galaxy, and environment around the source. Disentangling these components is a non-trivial problem as each component is not well known. Expected variations in DM from the IGM along an FRB path due to collapsed systems (e.g. halos) could be between $\sim$180-400\;pc\;cm$^{-3}$ at redshift unity \citep{mcqu14a,zhou14a}: much greater than uncertainties in the relation\footnote{$z \simeq {\rm DM}_{\rm exc}/(1200$ pc\;cm$^{-3} )$, (see \citet{ioka03a,petr16a})} conventionally used to estimate FRB redshifts \citep{petr16a}. Host contributions to total DM have been investigated (see, e.g., \citet{xu15a,yang17a}), and could potentially lie in the thousands of pc\;cm$^{-3}$, but the need for statically significant numbers of FRBs has been noted. Redshift errors associated with MW electron distribution models are not generally quoted; in addition, analysis of pulsars with parallax-determined distances has demonstrated the NE2001 model's potential to both overestimate and underestimate DM contributions by the MW, thus under/overestimating ${\rm DM}_{\rm exc}$, the most extreme result of which could be the misclassification of an FRB as another transient phenomenon, or vice-versa \citep{kean16a}. All of these factors hinder our ability to pinpoint the true redshifts and distances of poorly localised FRBs.

The purposes of this paper are as follows:
\begin{enumerate}
\item To apply physically realistic scenarios to FRB DM-redshift analysis by: (a) considering host contributions to DM; (b) following \citet{mcqu14a}, accounting for the cosmic variance of the IGM during analysis; (c) following \citet{zhou14a}, allowing FRB progenitor evolution across cosmic time to be accounted for, thereby allowing for more realistic FRB redshifts to be obtained from their DMs.
\item To provide methods for more rigorous analysis of FRB redshifts and their uncertainties, and by doing so improve FRB localisation potential via their DMs.
\item To illustrate the effects of different progenitor scenarios on FRB probability distributions, which may inform to what level DM components may be disentangled for large populations and potentially provide valuable information for indirect determination of FRB hosts.
\end{enumerate}

In Sect.~\ref{sect:2} we review the individual components which contribute to the dispersion measures of FRBs, and their relevance. In Sect.~\ref{sect:3} we combine these components and present a framework which can be used to place constraints on the redshifts of FRBs via their DM/redshift probability distribution functions (hereafter PDFs). Section~\ref{sect:4} discusses the implications of our findings. The methods presented in this paper can be extended to incorporate more realistic distributions of DM once they emerge.

Unless otherwise stated, in this paper we adopt a spatially flat $\Lambda$CDM cosmology model with the cosmological parameters fixed to the best-fit values of \citep{Planck_parameters}: $\Omega_{\rm b}=0.048$, $\Omega_{\rm m}=0.309$, $\Omega_{\Lambda}=0.691$, $n_{\rm s} = 0.9608$, $\sigma_{8}=0.815$, $H_0=67.3$.

\section{Contributions to FRB Dispersion Measure}\label{sect:2}
The observed DM of a single extragalactic FRB may be deconstructed into the following components:

      \begin{equation}
         {\rm DM}_{\rm obs} = {\rm DM}_{\rm MW}+{\rm DM}_{\rm IGM}+{\rm DM}_{\rm Host}+{\rm DM}_{\rm Local}
      \label{Eq1}
      \end{equation}
      
which represent contributions from free electrons along the light path of the burst due to the MW, IGM, its host galaxy, and its local environment (dependent on its progenitor model) respectively \citep{macq15a}. The excess extragalactic contribution to ${\rm DM}_{\rm obs}$ (i.e. ${\rm DM}_{\rm IGM}+{\rm DM}_{\rm Host}+{\rm DM}_{\rm Local}$) is referred to in this work as ${\rm DM}_{\rm exc}$.

\subsection{The Milky Way}\label{sect:2.1}
Typically ${\rm DM}_{\rm MW}$ is calculated using an electron distribution model such as NE2001 \citep{cord02a} or YMW \citep{yao17a} and subtracted from ${\rm DM}_{\rm obs}$. If completely accurate this leaves any remaining DM solely due to extragalactic electrons. Uncertainty in the NE2001 model has been discussed before in \citep{kean16a}, where it was noted that in its most extreme manifestation, error in the NE2001 model could lead to mis-classification of FRBs as Galactic repeating radio transients (hereafter RRATs), or vice-versa. In particular \citet{kean16a} highlights a source, J1354+24, as having a 20-40\;\% probability of being an FRB wrongly labelled an RRAT. These uncertainties will naturally cause difficulties while attempting to disentangle the true values of FRB DM components. For the purposes of this paper we discard the ${\rm DM}_{\rm MW}$ portion from subsequent analysis (see Sect.~\ref{sect:4} for a discussion).

\subsection{The Intergalactic Medium}

   \begin{figure}
   \centering
   \includegraphics[width=9cm]{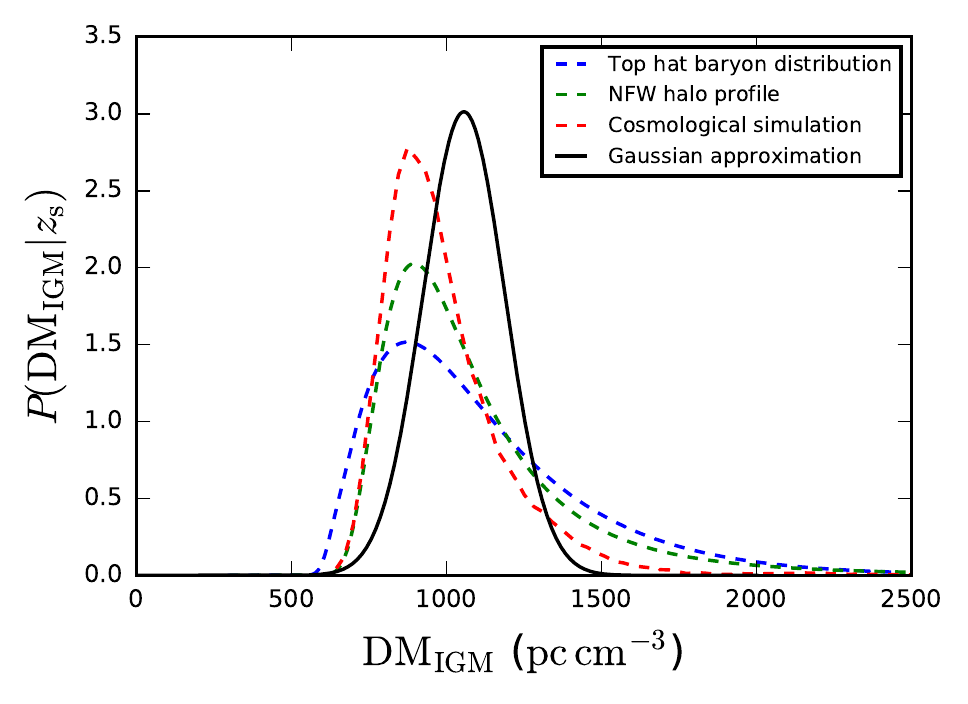}
      \caption{Comparison between ${\rm DM_{IGM}}$ probability distribution functions ($P({\rm DM}_{\rm IGM})$) for FRBs originating at $z=1$ for our Gaussian approximation taking into account sightline-to-sightline variance due to collapsed systems, and the more complex baryonic halo models (tracing a top hat function, NFW profiles, and cosmological simulations) from \citet{mcqu14a} (via. private communication). All profiles are normalised to $\int P({\rm DM}|z_{\rm s}=1)\;\der\, {\rm DM} = 1000$.}
         \label{FigIGM}
   \end{figure}

Approximately 40\% of all baryonic matter has been accounted for within galaxies, X-ray corona, and warm intergalactic matter \citep{mcqu14a}. More recently, matter was also detected in large-scale ionised gaseous filaments through use of the thermal Sunyaev-Zel'dovich effect \citep{degr17a,Tanimura17}. An extragalactic FRB will incur part of its DM budget during propagation through the ionised electron content of the IGM, thus serving as a potential probe of this and any further undetected baryonic matter along its travel path \citep{mcqu14a}. Standard analysis of ${\rm DM}_{\rm IGM}$ assumes a completely ionised population of baryonic matter, homogeneously distributed following \citet{ioka03a}, which simplifies the full relation between redshift and FRB dispersion measure to $z={\rm DM}_{\rm extragalactic}/1200$ pc cm$^{-3}$ \citep{petr16a}. Under these assumptions the theoretical relationship between expected DM and redshift is recovered to 2\% accuracy for FRBs originating closer than $z=2$. Sightline variations in the IGM due to halos along a propogation path, however, contribute a much larger uncertainty which in the future must be accounted for in order to obtain more realistic ${\rm DM}_{\rm IGM}$ constraints \citep{mcqu14a,petr16a}. In fact, whether any missing matter lies within galactic halos or further out could possibly be informed by probability distributions of FRB DMs, and so more complex models of the distribution of the IGM have previously been calculated analytically and via simulation, taking into account the sightline-to-sightline variance and different halo gas profiles \citep{mcqu14a}.

As the IGM contribution to FRB DMs becomes increasingly important with increasing redshift \citep{mcqu14a,ioka03a} it is therefore critical that models hoping to extract information from FRB DM probabilities take IGM behaviour into account as well as possible. Following \citet{mcqu14a,mo02a}, for FRB sources at redshift $z_{\rm s}$ in our example model we use a Gaussian approximation:

      \begin{equation}
       P({\rm DM}_{\rm IGM}|\;z_{\rm s})=\frac{1}{\sqrt{2\pi}\sigma_{\rm DM}}\exp\left(-\frac{({\rm DM}-{\rm DM}_{\rm IGM}(z_{\rm s}))^2}{2\sigma_{\rm DM}^2}\right)
      \end{equation}
      
with mean DM:

      \begin{equation}
        {\rm DM}_{\rm IGM}(z_{\rm s}) = \int^{\chi(z_{\rm s})}_{0}\der \chi\;\frac{n_{\rm e}(z)}{(1+z)^2}, \label{eq:DMz}
      \end{equation}
      
where the $n_{\rm e}(z)$ is the 3-dimensional electron number density. The variance (equal to the sightline-to-sightline DM scatter) for the probability distribution function of ${\rm DM_{IGM}}$ is:
      \begin{equation}
        \sigma_{\rm DM}^2\left({\rm DM},z_{\rm s} \right) = \int^{\chi_{\rm s}}_{0}\der \chi (1+z_{\rm s})^{2}\bar{n}^{2}_{\rm e}(0)\int\frac{\der^{2}k_{\perp}}{(2\pi)^{2}}P_{\rm e}(k_{\perp},z_{\rm s}),
      \end{equation}
      
where $\bar{n}_{\rm e}(0)$ is the mean electron density at redshift $0$, and the comoving distance at redshift $z$ is  $\der\chi = c\der z/H(z)$. The matter power spectrum at $z$ for wavenumber $k$ is $P_{\rm e}(k,z)$, where $k_{\perp}$ is the perpendicular component. We compute the matter power spectrum from public code \textsc{camb}\footnote{\url{https://camb.info}}, with {\it Planck} best-fitting cosmological parameters \citep{Planck_parameters}.

Taking into account this cosmic variance allows for more realistic IGM models, and more complex profiles \citep[such as gas profiles as simulated in][]{mcqu14a} may be substituted in the future. Figure~\ref{FigIGM} shows a comparison of our Gaussian approximation with McQuinn's models at redshift $1$. For more detailed calculations, see Appendix~\ref{sec:dispersion}.

\subsection{The Host Galaxy}

The second source of extragalactic electrons which influences dispersion are FRB host galaxies. ${\rm DM}_{\rm host}$ may vary drastically depending on the assumed FRB progenitor model, being intrinsically subject to: (1) the FRB's host galaxy type, and (2) its location therein.

The host galaxy contribution to DM has previously been explored for GRBs and found to contribute up to 10$^5$ pc\;cm$^{-3}$ if GRBs originate in star-forming regions \citep{ioka03a}. \citet{yang17a} estimate the mean host contribution for 21 observed FRBs to be $\sim 270$\;pc\;cm$^{-3}$. ${\rm DM}_{\rm host}$ distributions have also been simulated and found to contribute up to a few thousands or tens of pc\;cm$^{-3}$ for rest-frame spiral or elliptical galaxies respectively, when taken as a scaled versions of the MW as modelled by NE2001 \citep{xu15a}. As ${\rm DM}_{\rm host}$ scales with source redshift as $1/(1+z)$ \citep{macq15a}, this may or may not contribute a significant amount to the total ${\rm DM}_{\rm obs}$. However it is still important to account for this portion during comprehensive DM-distance analysis, especially at low redshifts.

By simulating many FRBs within a given model, probability distributions may be built for FRB  progenitor scenarios in the host rest frame. For sources at a given redshift $z_{\rm s}$, the DM PDF then becomes:

\begin{equation}
P({\rm DM_{host}}|\;z=z_{\rm s})= (1+z_{\rm s})P((1+z_{\rm s}){\rm DM_{host}}|z=z_{0})
\end{equation}

In the remainder of this section we discuss our simple models for different galaxy classes that FRBs may originate within, and how they may be populated according to various progenitor scenarios.
   
\subsubsection{Galaxy Classes}\label{sect:2.3.1}

For simplicity, we group our models into two classes of host galaxy: spiral and elliptical. In practice, we create host galaxy electron distribution models for each type, populate the models with FRBs, and by integrating electron densities along lines of sight obtain DM probability densities at the rest-frame of the hosts. These are then collected into redshift-scalable PDFs. Below we discuss our method for modelling both classes.

\begin{table}
\caption[]{Spiral galaxy electron distribution model parameters (see \citet{gome01a,schn12a,kalb08a}).}
\label{table:GBCa}
$$
\begin{array}{p{0.5\linewidth}l}
\hline
\noalign{\smallskip}
Parameter & {\rm Value}\\
\noalign{\smallskip}
\hline
$h_{r,1}$ & 30.4 \mathrm{\;kpc}\\
$h_{r,2}$ & 1.5 \mathrm{\;kpc}\\
$h_{z,1}$ & 1.07 \mathrm{\;kpc}\\
$h_{z,2}$ & 0.050 \mathrm{\;kpc}\\
$n_1$ & 2.03\times10^{-2}\mathrm{\;cm}^{-3}\\
$n_2$ & 0.71\times10^{-2}\mathrm{\;cm}^{-3}\\
$r_f$ & 17.5\mathrm{\;kpc}\\
$r_n$ & 3.15\mathrm{\;kpc}\\
\noalign{\smallskip}
\hline
\end{array}
$$
\end{table}

\begin{figure*}
\centering
\includegraphics[width=18cm]{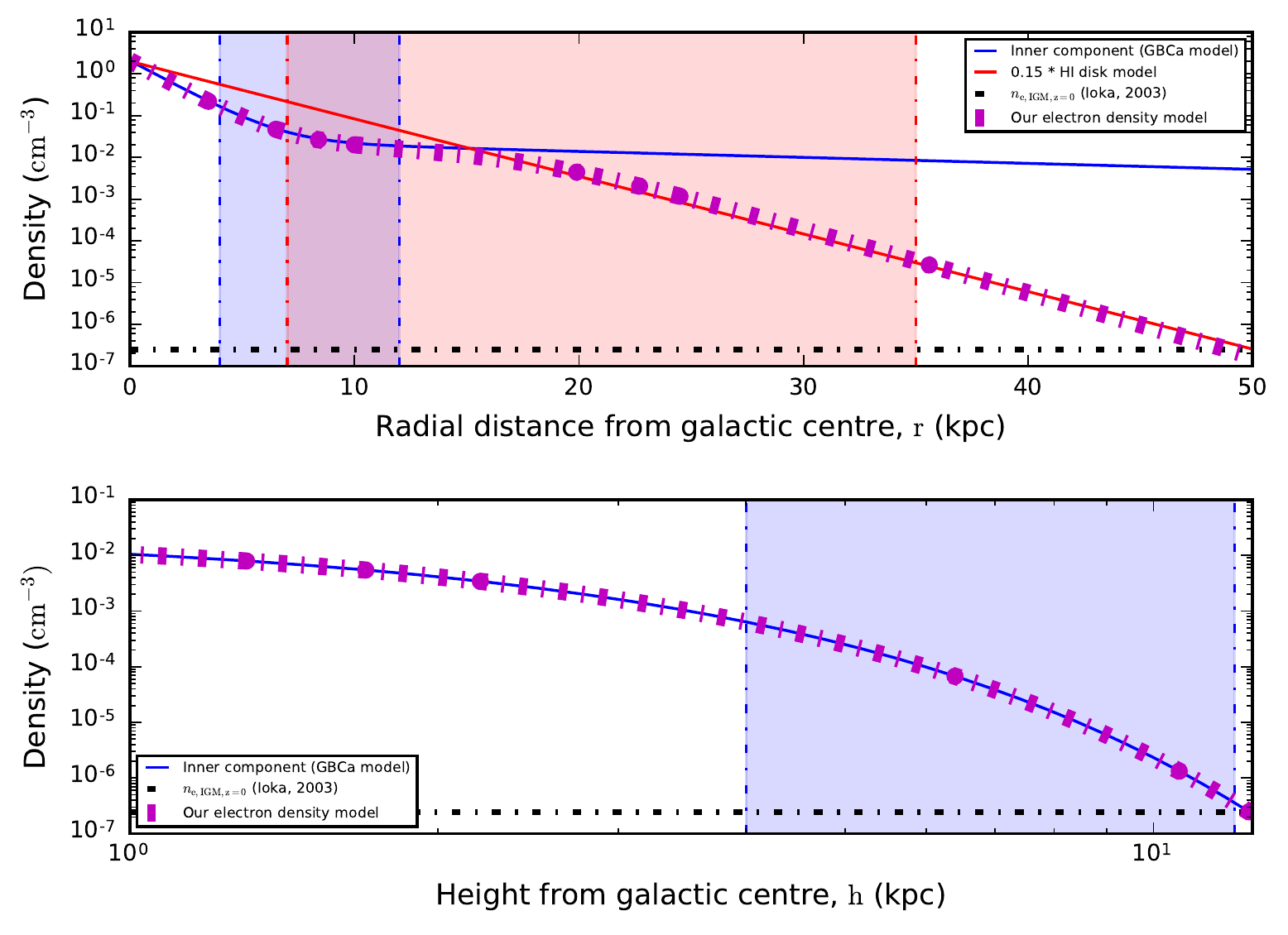}
\caption{Radial and height cross sections of our spiral electron density model (magenta, dashed), consisting of an inner component tracing the MW Galactic disk \citep{gome01a,schn12a} (blue line), and an outer component approximately following the MW HI disk \citep{kalb08a} with an ionisation fraction $\sim15$\% (red line). Shaded (blue/red) regions highlight the areas in which the inner/outer profiles are known to be well-constrained respectively. Following \citet{ioka03a}, an approximate electron density for the IGM at redshift zero, $n_{\rm e, IGM, z}$, is also plotted (black, dashed).}
\label{FigSpiral}
\end{figure*}

~\\
\textit{Spiral Galaxies}\\
\\
Previously, \citet{xu15a} scaled the NE2001 model \citep{cord02a} (which details fine structure of the MW) to represent typical spiral galaxies for FRB analysis. However, simpler electron distribution models have also been found to reproduce DMs to pulsars with known parallax distances to within a factor 1.5-2 \citep{schn12a}. Such a model is sufficient to capture the main properties of spiral galaxies for statistical studies. For demonstrative purposes we combine two simple components for our model. The first component describes the inner Milky Way Galactic disk, where the majority of stellar mass is concentrated, using a sum of two ellipsoids made of decaying exponentials. Denoted the GBCa model by \citet{schn12a}, this model was developed using 109 pulsars with distances determined independently of their DMs \citep{gome01a}. The GBCa model is well constrained within 4-12\;kpc of the Galactic center, and within the central 4\;kpc we make the assumption it is acceptable in the present context.

Further out in spiral galaxies where stellar mass is more difficult to trace we use neutral hydrogen (HI) as a proxy. The outer component of our model is based on a single decaying exponential, approximately describing the average mid-plane volume density of the Milky Way's HI disk between 7-35\;kpc \citep{kalb08a}.

These two components are combined into a smoothly varying function:

\begin{equation}
n_{\rm e,spiral}(r,h) = \frac{1}{1+\exp((r-r_f)/r_n)}\sum^{2}_{i=1}n_{i}\frac{\exp(-r/h_{r,i})}{\exp(-r_{\odot}/h_{r,i})}\exp\left(\frac{-h}{h_{z,i}}\right)
\end{equation}

with cylindrical coordinates $r$ and $h$ denoting radial distance from the Galactic center and height above the Galactic plane respectively. All relevant parameters are provided in Table \ref{table:GBCa}.

In order to reconcile the two components, the scaling of the original HI function from \citet{kalb08a} requires an ionization fraction of $\sim15$\%. In addition, at a radius $\sim50$\;kpc $n_{\rm e,spiral}(r,h)$ reaches a value consistent with the electron density of the IGM at redshift zero, as estimated by \citet{ioka03a}. The model and its components may be seen in Figure \ref{FigSpiral}.

~\\
\textit{Elliptical Galaxies}\\
\\
Elliptical galaxies may be decomposed into four separate parts: a central black hole, stars, diffuse hot gas, and dark matter \citep{mamo05a}. To model a $5\times10^{10}\;{\rm M_{\odot}}$ elliptical galaxy we assume ionised electrons trace a hot gas profile following \citet{mamo05a}, which we write as:

\begin{equation}
n_{\rm elliptical}(r)=\rho_0 \left(1+\left(\frac{r}{r_{\rm c}}\right)^2\right)^{-3\beta_{\rm g}/2}
\end{equation}

with central galaxy density $\rho_0 \simeq 6\times10^7\;{\rm M_{\odot}\;kpc^{-3}}$, $\beta_{\rm g}=0.5$ and where $r_{\rm c}$ is related to the effective radius of the galaxy (radius of half-projected light) $R_{\rm e}$ by

\begin{equation}
r_{\rm c} \simeq \frac{R_{\rm e}}{q}
\end{equation}

where $q=10$. Our ionisation fraction (0.04\%) and $R_{\rm e}$ (3.2\;kpc) were chosen so the model approached the $z=0$ IGM electron density at $\sim 70kpc$. For a full derivation and assumptions related to baryonic fractions and central black hole masses in elliptical galaxies which were adopted by the model, see \citet{mamo05a}.
   
\subsubsection{FRB Distributions Within Galaxies}

   \begin{figure*}
   \centering
   \includegraphics[width=18cm]{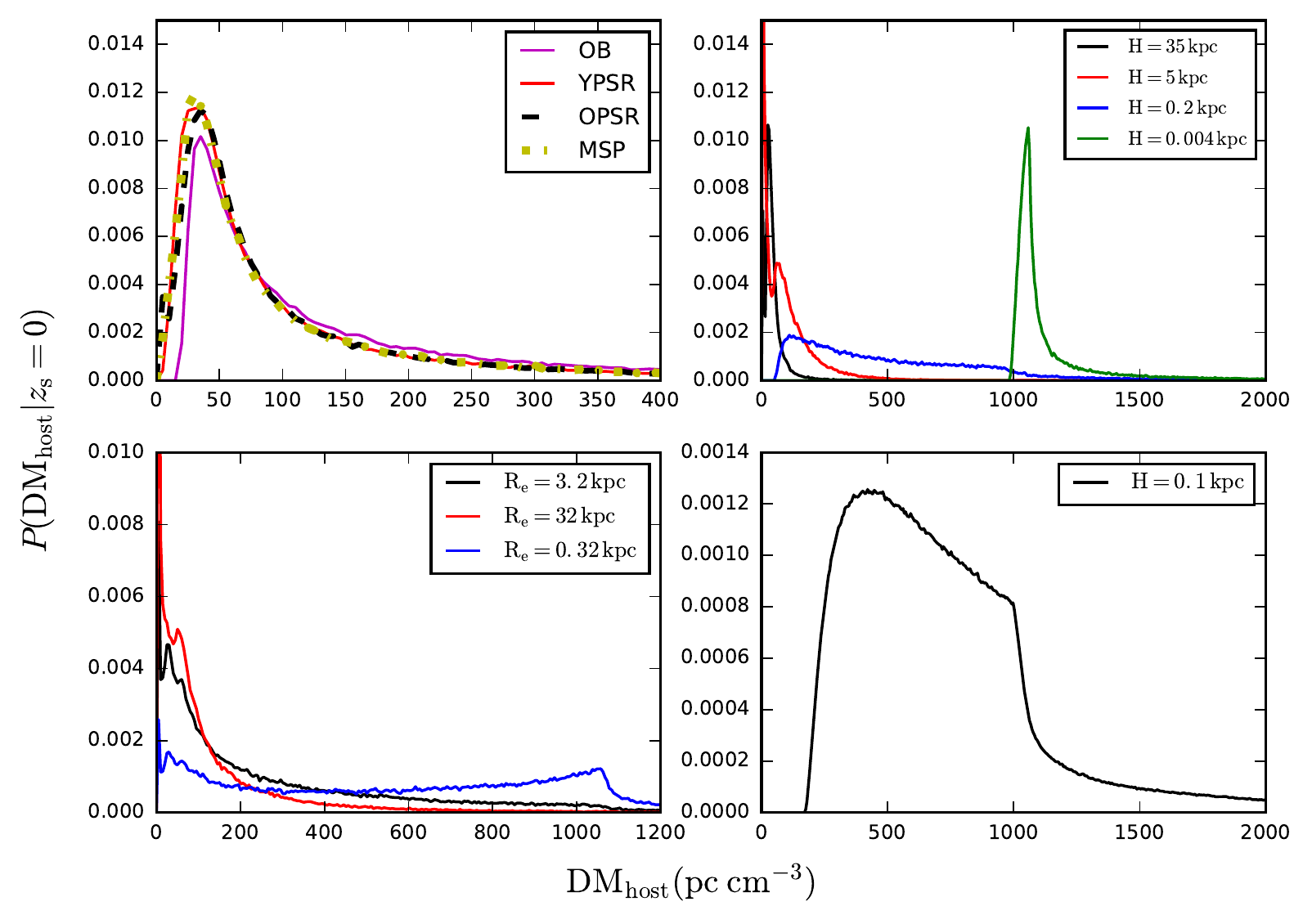}
      \caption{Model rest-frame DM PDFs ($P({\rm DM}_{\rm host})$) for FRBs originating from different distributions within different galaxy types. Top left to bottom right: Stellar distributions (OB stars, young/old pulsars, millisecond pulsars) in spiral galaxies; homogeneous distributions (within spheres of radius ${\rm H}$ kpc) in spiral galaxies; elliptical distributions (confined to multiples of the effective (half light) radius, ${\rm R_e}=3.2\,{\rm kpc}$) in elliptical galaxies; homogeneous distributions (within spheres of radius ${\rm H}$\,kpc) in elliptical galaxies of effective radius 3.2\,kpc.}
         \label{FigHost}
   \end{figure*}
   
In this section we demonstrate the effects of different FRB distributions within our galaxy types on their DMs. Our distributions are either phenomenological in nature, tracing stellar populations such as, e.g., OB stars and various ages/types of pulsars (denoted Stellar Distributions), or are homogeneously generated to appear within given radii from the galactic center (denoted Homogeneous Distributions). The former models imply mostly FRB-neutron star (hereafter NS) associations; the latter models imply association with AGNs or galactic center halos.

FRBs were generated to populate both spiral and elliptical galaxies according to the electron distributions considered in Section~\ref{sect:2.3.1}. DMs were calculated for outside observers viewing from random orientations. The resulting probability distribution functions are shown in Fig. \ref{FigHost}.

\begin{table}
\caption[]{Parameters used in pulsar distribution models \citep{yusi04a,sun04a,maiz01a}.}
\label{table:yusi}
$$
\begin{array}{p{0.5\linewidth}l}
\hline
\noalign{\smallskip}
Parameter & {\rm Value}\\
\noalign{\smallskip}
\hline
$\sigma$ & 7.5\;{\rm kpc}\\
$r_{\odot}$ & 8.5\;{\rm kpc}\\
$r_1$ & 0.55\pm0.10\;{\rm kpc}\\
$A_{\rm PSR}$ & 1.64\pm0.11\\
$a$ & 1.64\pm0.11\\
$b$ & 4.01\pm0.24\\
$h_{\rm OB}$ & 63\;{\rm pc}\\
$h_0$ & 49\times10^{-3}\;{\rm kpc}\\
$\sigma_{\rm PSR}$ & 282\;{\rm km\;s^{-1}}\\
$h_{\rm OPSR}$ & 0.4\;{\rm kpc}\\
$t\,(\rm YPSR)$ & 2\;{\rm Myr}\\
$t\,(\rm OPSR)$ & 1\;{\rm Gyr}\\
$h_{\rm MSP}$ & 0.5\;{\rm kpc}\\
\noalign{\smallskip}
\hline
\end{array}
$$
\end{table}

~\\
\textit{Stellar Distributions (Spiral Galaxies)}\\
\\
To model the distributions of FRBs within spiral galaxies, we consider different scenarios based on theoretical FRB progenitors. FRBs as outbursts from flaring OB stars \citep{loeb14a} would trace young stellar populations in a galaxy; FRBs as outbursts from magnetars \citep{pen15a}, pulsar supergiant pulses \citep{cord16a}, NS-NS mergers \citep{tota13a}, pulsar companions \citep{mott14a}, or collapsing NS \citep{falc14a} should follow NS distributions of various ages. Our chosen models were two-component (radial and height) distributions: for millisecond pulsars (hereafter MSPs), old and young pulsars (hereafter OPSRS, YPSRs; valid for characteristic ages greater/less than 8\;Myrs respectively), and for OB stars.

In order to account for high young stellar population densities and large populations of MSPs close to the Galactic center, these objects were modeled using Gaussian radial distributions after \citet{lori13a}:

\begin{equation}
R_{\rm OB}(r)=R_{\rm MSP}(r)=\frac{1}{\sqrt{2\pi}\sigma}\exp{\left(-\frac{r^2}{2\sigma^2}\right)}
\end{equation}

To maintain consistency with the observed pulsar dearth in the central Galactic kpc \citep{macq15a}, OPSR and YPSR radial distributions were modeled with a gamma function after \citet{yusi04a}:

\begin{equation}
R_{\rm YPSR}(r)=R_{\rm OPSR}(r)=A_{\rm PSR}\left(\frac{X}{X_{\odot}}\right)^a\exp\left(-b\left(\frac{X-X_{\odot}}{X_{\odot}}\right)\right)
\end{equation}

where $X=r+r_1$ and $X_\odot=r_\odot+r_1$.\\
Height distributions for OB stars, YPSRs, and OPSRs were modelled following \citet{maiz01a,sun04a}:

\begin{equation}
H_{\rm OB}(h)=\frac{1}{\sqrt{2\pi}h_{\rm OB}}\exp\left(-\frac{h^2}{2h_{\rm OB}^2}\right)
\end{equation}

\begin{equation}
H_{\rm YPSR}(h)=\frac{2}{\sqrt{2\pi}h_{\rm PSR}}\exp{\left(-\frac{h^2}{2h_{\rm PSR}^2}\right)}
\end{equation}

\begin{equation}
H_{\rm OPSR}(h)=\frac{2}{\sqrt{2\pi}h_{\rm PSR}}\exp{\left(-\frac{h^2}{2h_{\rm PSR}^2}\right)}+\frac{1}{h_{\rm OPSR}}\exp{\left(\frac{\mid h \mid}{h_{\rm OPSR}}\right)}
\end{equation}

with scale height: $h_{\rm PSR}=h_0+\sigma_{\rm PSR} t$.

The height distribution for MSPs was modelled again after \citet{lori13a}:

\begin{equation}
H_{\rm MSP}(h)=\frac{1}{h_{\rm MSP}}\exp{\left(-\frac{h}{h_{\rm MSP}}\right)}
\end{equation}
Parameters are provided in Table \ref{table:yusi}.

~\\
\textit{Elliptical Distributions (Elliptical Galaxies)}\\
\\
For elliptical galaxy distributions, we consider the scenario where FRBs progenitors trace the stellar mass distribution, and thus the ionised electron content, of the galaxy. In this case the host galaxy was populated with sources generated following an identical profile to that of the host's $n_{\rm e}$ content, normalised to the number of progenitors chosen for the simulation.

~\\
\textit{Homogeneous Distributions (Spirals and Ellipticals)}\\
\\
Additionally, we create homogeneous FRB distributions around the galactic centers of both spiral and elliptical galaxies by populating FRBs randomly within spheres of variable radius, such as could be expected from central galactic halos. For smaller limiting radii, this scenario could represent FRB mechanisms connected to AGN, for example interactions between relativistic jets and plasma \citep{rome16a,viey17a}. Larger limiting radii could be consistent with progenitors located in more extended galactic halos.

\subsection{The Burst Environment}

The final contribution to the DM of an FRB, ${\rm DM}_{\rm local}$, is the local electron environment of the burst itself, which varies wildly depending on the FRB progenitor model. An extreme example of this is considered by \citet{loeb14a}, who propose that flares from main sequence stars, albeit from Galactic sources as close as 1 kpc, could produce FRBs which acquire their ${\rm DM_{exc}}$ while propagating through their host's coronal plasma. For the purposes of this paper we assume extragalactic origins for FRBs and idealistically discard the burst component.

\section{The  Excess Electron Model}\label{sect:3}

\begin{figure*}
\centering
\includegraphics[width=18cm]{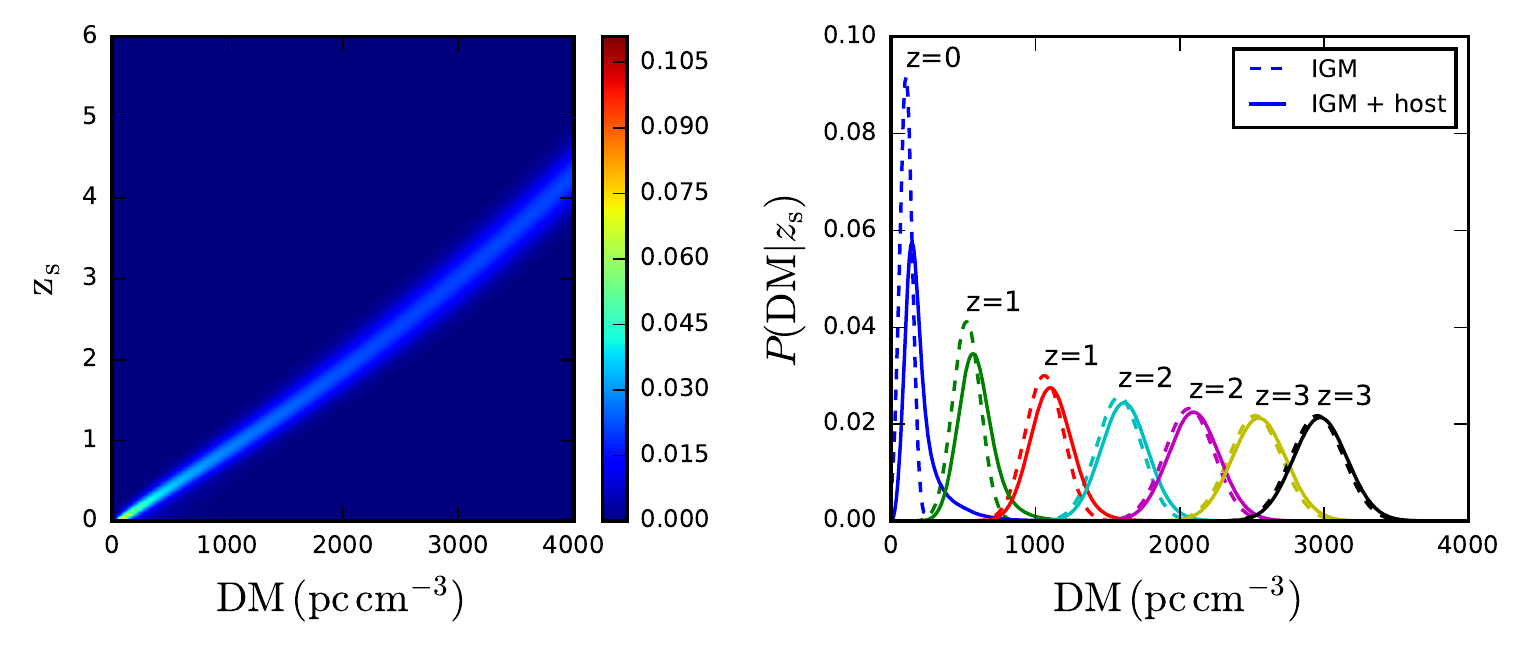}
\caption{Excess electron model for a distribution of young pulsars in spiral galaxies. Left: The probability distribution of FRBs as a function of DM and source redshift ($P({\rm DM}|z{\rm _{s}})$). Right: Projections of the model along different source redshifts (solid lines) plotted against $P({\rm DM_{IGM}}|z{\rm _{s}})$, the DM probability distribution function of the intergalactic medium (dashed lines). All curves are normalised to unity.}
\label{FigHostRedshiftDependence}
\end{figure*}

In this section we combine the extragalactic components which contribute to FRB DMs (excluding the local burst environment) for our chosen progenitor scenarios in order to create a model describing the relationship between FRB dispersion measures and redshifts. The method is applicable not only to our IGM and progenitor models, but also to any future models deemed more physically accurate as new information arises from future FRB discoveries. In subsection 3.1 we describe our model, then in subsection 3.2 we invert it and discuss advantages of doing so.

\subsection{P({\rm DM}|z) methodology and results}

Chosen models for the IGM and host galaxy may be convolved to create probability distributions for the total ${\rm DM}_{\rm exc}={\rm DM}_{\rm IGM}+{\rm DM}_{\rm host}$ (excluding ${\rm DM}_{\rm local}$):

\begin{equation}
\label{eq:conv}
P({\rm DM}_{\rm exc}|\;z_{\rm s}) = P({\rm DM}_{\rm IGM}|\;z_{\rm s}) \ast P({\rm DM}_{\rm host}|\;z_{\rm s})
\end{equation}

which describe the likelihood of an FRB of a particular progenitor at a given redshift of having a specific DM. We will refer to this as the excess electron model.

\begin{figure*}
\centering
\includegraphics[width=18cm]{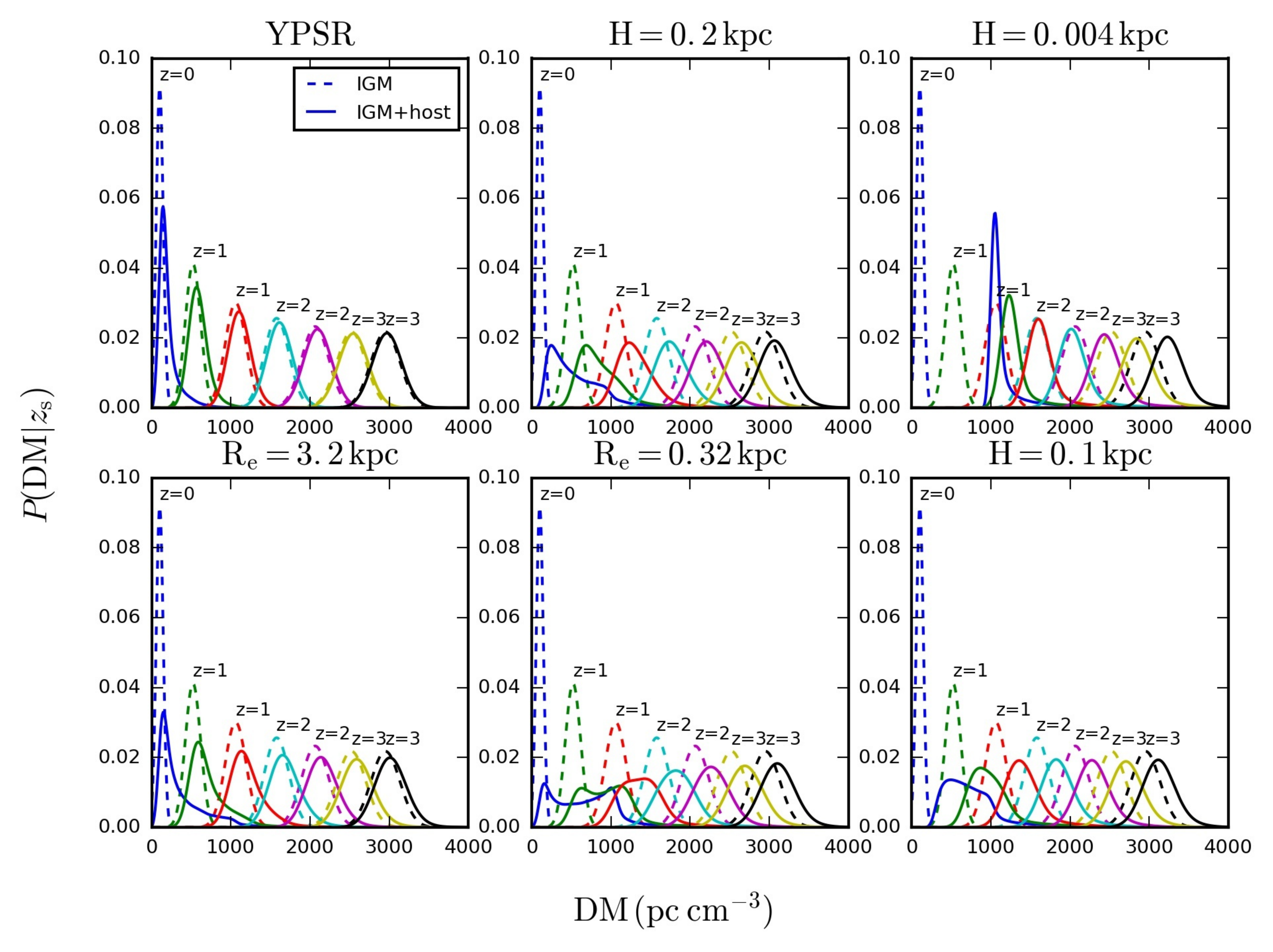}
\caption{Probability distributions illustrating the relative contributions of host galaxy models in spiral galaxies (top panel) and elliptical galaxies (bottom panel) to total DM of FRBs out to redshift 3. Top panel (left to right): young pulsars, random distributions within spheres of radius $H$ about the galactic center. Bottom panel (left to right): elliptically distributed FRBs confined to multiples of the galaxy half light radius (3.2\;kpc), random distribution, as above. Solid lines show probability distributions where both IGM and host galaxy contributions have been considered. Dashed lines show probability distributions where the host galaxy contribution is neglected.}
\label{FigHostRedshiftComp}
\end{figure*}

An example excess electron model, created using the YPSR in spiral galaxies progenitor model is shown in Fig. \ref{FigHostRedshiftDependence} (left). This model spans redshift and DM ranges 0-6 and DMs 0-4000\;pc\;cm$^{-3}$ respectively, and is normalised to ensure $\int P({\rm DM}_{\rm exc}|z_{\rm s})\der\, {\rm DM}=1$. Figure \ref{FigHostRedshiftDependence} (right) demonstrates how probability of detecting an FRB with a particular DM changes with increasing source redshift $z_{\rm s}$ using horizontal projections of the model. As can be seen, at low source redshifts, the host galaxy contribution plays an influential role in increasing the possible FRB DM range to large DMs. This influence diminishes with increasing $z$. Figure \ref{FigHostRedshiftComp} compares this effect for several of our host models for spiral (young pulsars and homogeneous distributions close to galactic centers) and elliptical (elliptical distributions confined to 0.1, and 1 times the half light radius; and a homogeneous distribution close to the galactic center) galaxies. As can be seen, the DM probability distributions vary noticeably at low z. For large populations of FRBs with accurately known redshifts, this could be a potentially useful tool for identifying hosts (see Sect.~\ref{sect:4}). 

\subsection{P(z|{\rm DM}) methodology and results}

\begin{figure}
\centering
\includegraphics[width=9cm]{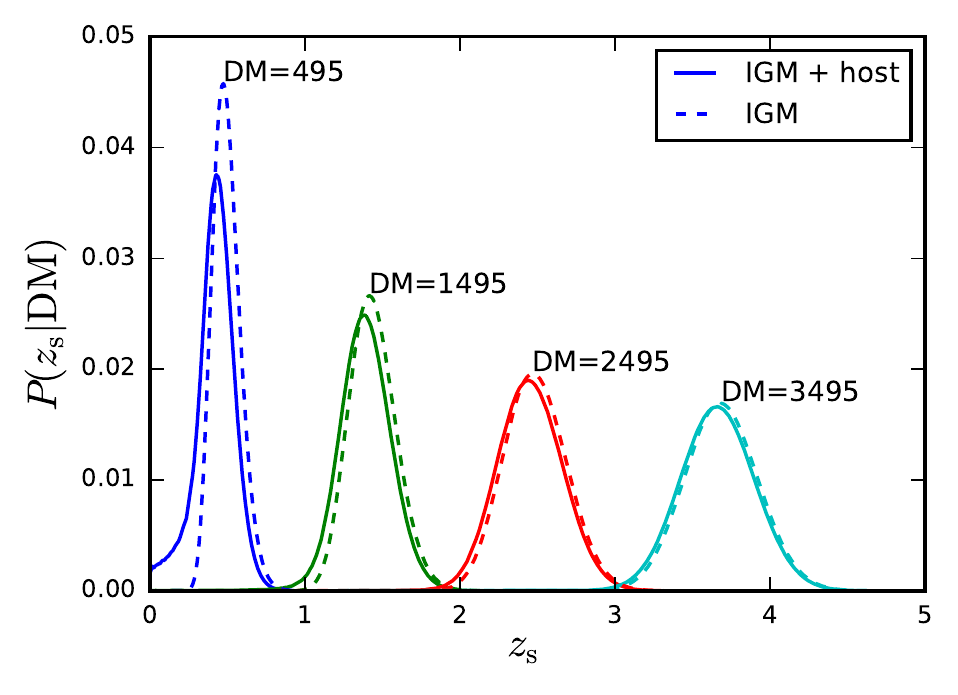}
\caption{Redshift probability distributions for populations of FRBs with different DMs drawn from a young pulsars in spiral galaxies excess electron model (solid lines) against distributions only accounting for the intergalactic medium (dashed lines). All curves are normalised to $\int P(z_{\rm s}|{\rm DM})\,dz=1$.}
\label{FigDMcuts}
\end{figure}

An alternative use for our model arises from its inversion. According to Bayes' theorem,

\begin{equation}
\label{eq:bayes}
P(z_{\rm s}|\;{\rm DM}_{\rm exc})= \frac{P({\rm DM}_{\rm exc}|\;z_{\rm s})P(z_{\rm s})}{P({\rm DM}_{\rm exc})},
\end{equation}

where the $P(z_{\rm s})$ is the assumed prior redshift distribution of FRBs, and $P({\rm DM}_{\rm exc})$ is the probability of the DM excess. When normalised to $\int P(z_{\rm s}|\;{\rm DM}_{\rm host})\,{\rm d}z=1$, projections of the model demonstrate how host models affect redshift probabilities for populations of FRBs with given ${\rm DM}_{\rm exc}$. This is demonstrated in Fig. \ref{FigDMcuts} using a young pulsars in spiral galaxies progenitor model. By comparing the model to probabilities generated neglecting host galaxy contributions, we demonstrate their lessening importance for FRBs when statistically analysing FRB populations with large DMs. This result is unsurprising due to the increased probability that a high-DM FRB will originate at a larger redshift, thus decreasing the importance of ${\rm DM_{host}}$. At lower dispersion measures, however, the ensemble effect of host galaxies becomes noticeable.

\begin{figure}
   \centering
   \includegraphics[width=9cm]{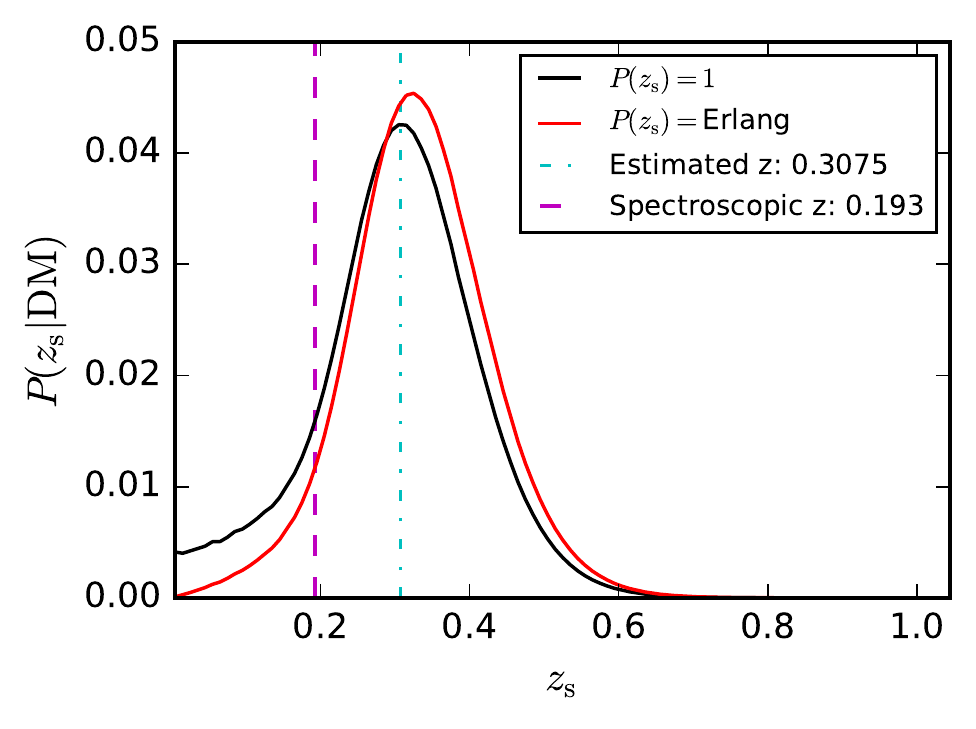}
      \caption{Redshift probability distributions normalised to $\int P(z_{\rm s}|{\rm DM})\,dz=1$ for repeating FRB 121102. Curves were drawn from an excess electron model assuming an FRB progenitor distribution of young pulsars in spiral galaxies. Black: assuming a homogeneous prior for FRB redshifts. Red: assuming an Erlang prior. Cyan, dash-dotted: the FRBcat estimate redshift (using standard practice of ${\rm DM}/(1200 {\rm pc\;cm^{-3}})$. Magenta, dashed: spectroscopic redshift of FRB 121102 host galaxy.  }
         \label{FigErlang}
   \end{figure}

Finally, inversion of the model allows assumptions to be made and tested regarding the redshift distribution of FRBs across cosmic time. Setting $P(z_{\rm s})=1$ assumes a homogeneous redshift distribution for FRBs, i.e. the probability of occurrence and detection is equal for all redshifts considered in the model. This may not be the case; redshift distributions will rely on the phenomenology of FRB progenitors, and have previously been considered to follow the Erlang distribution fit for GRBs: 

\begin{equation}
P(z_{\rm s};k,\gamma)=\gamma^kz_{\rm s}^{k-1}e^{-\gamma z_{\rm s}}/\Gamma(k)
\end{equation}

with $k=2$, $\gamma=1$ \citep{zhou14a}. Both cases have been included in our model. An Erlang prior noticeably affects the range of redshifts possible for detected FRBs (see Fig. \ref{FigErlang}).

\section{Discussion and Conclusion}\label{sect:4}

   \begin{figure*}
   \centering
   \includegraphics[width=18cm]{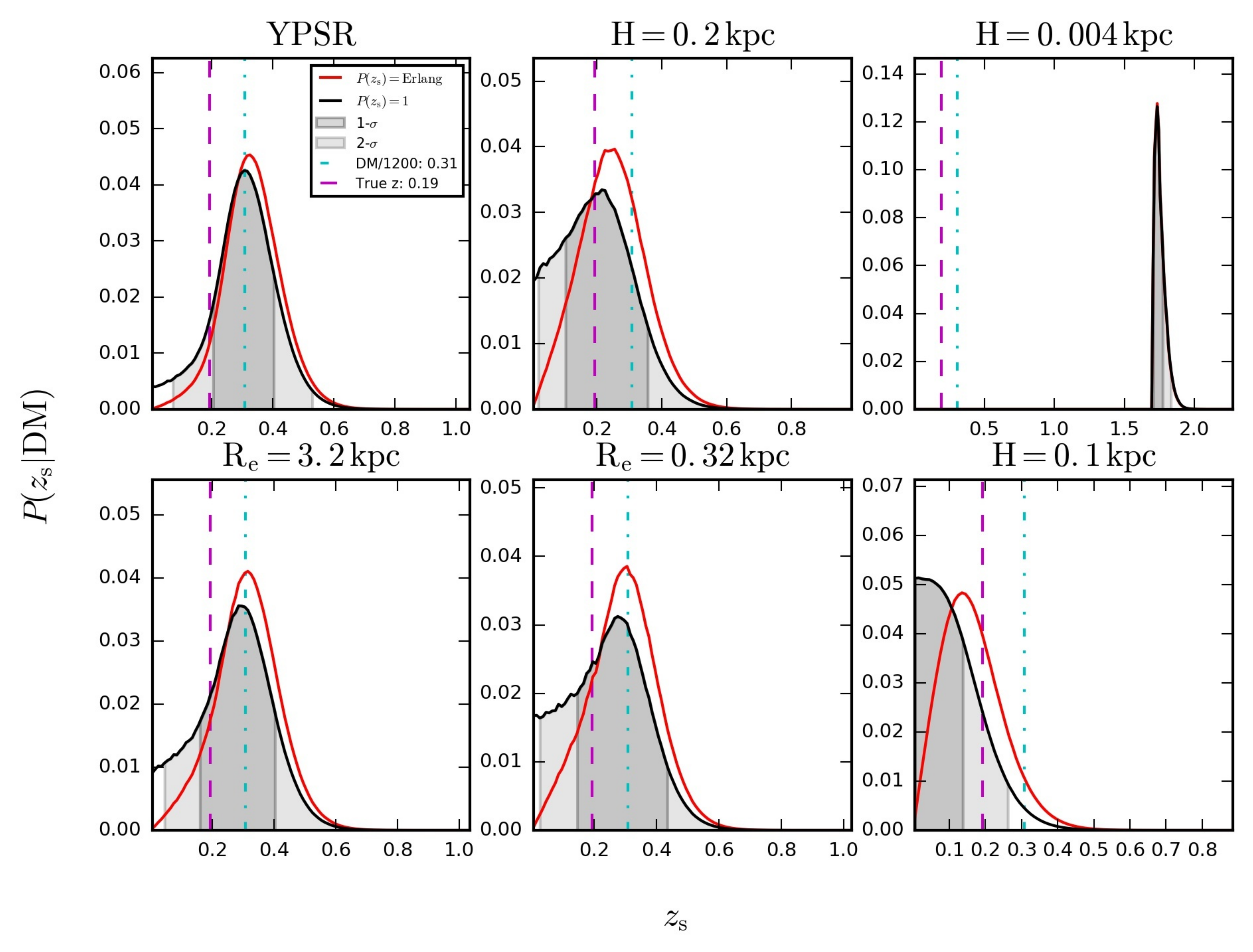}
      \caption{Redshift probability distributions for FRBs with DMs equal to FRB121102 according to models in spiral galaxies (top panel) and elliptical galaxies (bottom panel). Top panel (left to right): young pulsars, random distributions within spheres of radius $H$ about the galactic center. Bottom panel (left to right): elliptically distributed FRBs confined to multiples of the galaxy half light radius (3.2\;kpc), random distribution, as above. Curves assume underlying homogeneous ($P(z_{\rm s})=1$, black lines) and Erlang ($P(z_{\rm s})=P(z_{\rm s};k,\gamma)$, red lines) priors for FRB redshifts. For homogeneous distributions, shaded regions mark bounds drawn from 68\% and 99\% confidence intervals. Standard DM-derived (cyan) and spectroscopically determined (magenta) values for FRB121102's redshift are included.}
         \label{FigWMBComp}
   \end{figure*}

In previous sections we summarized the contributing components to FRB DMs and created a model describing the probabilistic relationship between FRB DM and redshift. The total observed DM will always be well-defined for FRBs, while accurate redshifts, relying on arcsecond (or better) localization, may not always be obtained. Therefore the aims of our model are threefold:
\begin{enumerate}
  \item To improve FRB DM-redshift analysis by considering:
    \begin{itemize}
      \item The variance in DM acquired during propagation through the IGM due to halos.
      \item The contribution of the host galaxy to an FRB's DM under different progenitor assumptions.
      \item The effects of underlying redshift distributions for FRBs.
    \end{itemize}
  \item To place constraints on redshifts of unlocalized FRBs which have been calculated via their dispersion measures, given the above considerations.
  \item To illustrate the effects of different progenitor scenarios and allow the feasibility of disentanglement of FRB components and the identification of FRB hosts/progenitors (given large enough FRB populations) to be explored.
\end{enumerate}

\subsection{Improved analysis of FRB {\rm DM} components}

For every progenitor scenario we find DM host distributions to have heavy tails (see Fig. \ref{FigHost}), allowing FRBs to be dispersed significantly (>\;400\;pc\;cm$^{-3}$, and in some cases in excess of \;700\;pc\;cm$^{-3}$) by the host galaxy in the rest frame. This is in agreement with previous studies of spiral galaxies \citep{xu15a,zhou14a}. Our simulations also predict this behaviour for elliptical galaxies, in disagreement with past literature \citep{xu15a}. This discrepancy most likely comes from different modelling methods. \citet{xu15a} scale a MW-sized elliptical galaxy (constructed by neglecting the thin disk and spiral arm components of the NE2001 model) using ${\rm H_\alpha}$ as a tracer of ionized gas. We note that our chosen models may not capture the origins of FRBs on an individual basis. FRB121102, the only localised example of an FRB to date, originates in a dwarf galaxy \citep{tend17a}: a scenario we have not considered here. Future self-consistent models would ideally also incorporate any DM contribution due to FRB local burst environments.

Regardless of galaxy model, analysis could be improved further by factoring in a more realistic host galaxy mass distribution rather than using a fixed mass for both spiral and elliptical galaxies. Given that our assumed masses lie closer to the top of the mass range, this would make the contribution of the host to FRB DMs even more negligible. As demonstrated in Figure \ref{FigHostRedshiftComp} the relative contribution of the host galaxy to ${\rm DM_{exc}}$ is most influential at low redshifts and becomes less relevant for sources at large $z_{\rm s}$. This effect due to time dilation makes ${\rm DM}_{\rm host}$ fall off by a factor $1/(1+z_{\rm s})$ \citep{macq15a}. It has previously been noted to inadequately suppress host galaxy contributions larger than $700$\;pc\;cm$^{-3}$ for FRBs with $z_{\rm s}<2$ \citep{zhou14a}. This phenomenon takes place regardless of host galaxy/progenitor model. In the cases of some of our models, for example homogeneously distributed FRBs originating close to galactic centers, the host noticeably affects the probable DM of a burst even out to $z_{\rm s}\sim 3$.

Finally, Fig. \ref{FigErlang} illustrates the noticeable effect of underlying FRB redshift priors on their potential redshifts. Only homogeneous and Erlang distributions have been considered here; further FRB detections and localisations may inform the choice of redshift distribution in the future. The choice of Erlang is motivated by GRB studies: the deficit in the number of observed events in the nearby Universe is presumably caused by the relative paucity of galaxies, whereas sensitivity-limited instruments become less likely to detect increasingly distant events.

\subsection{Constraints on redshifts for unlocalised FRBs}

Table \ref{FRBRedshiftTable} contains redshift values for every catalogued FRB to date derived using several progenitor scenarios, with bounds drawn from their 99\% confidence intervals. The table is not a comprehensive list of all progenitor possibilities mentioned in this work, focusing on a few significantly different scenarios: (1) neglecting a host galaxy, (2) FRBs arising from young pulsar populations in MW-like spiral galaxies, (3) from compact (radius=0.2\,kpc, 0.1\,kpc, respectively) spheres in MW-like spirals or $5\times10^{10}\;{\rm M_{\odot}}$ elliptical galaxies, and (4) from distributions following gas profiles in $5\times10^{10}\;{\rm M_{\odot}}$ ellipticals; with a homogeneous distribution for redshifts across cosmic time ($P(z_{\rm s})=1$).

These scenarios serve to highlight the difference that progenitors can make to the range of redshifts available to FRBs of particular DMs. Constraints on redshift may be easily obtained for an FRB with these or any of our other progenitor scenarios using the models and jupyter notebook provided online. In addition, users may substitute their own rest-frame host galaxy FRB DM probability distributions and redshift distributions. We propose that such techniques be considered for general use during future FRB analysis, so as to better constrain the redshifts of unlocalised FRBs.

Inverting the excess electron model to obtain $P(z_{\rm s}|{\rm DM})$ may allow for an innovative method for constraining hosts for unlocalized FRBs. This remainder could be fed back into the excess electron model, to order to obtain the likelihood that an FRB came from each galaxy in its field of observation, subject to progenitor assumptions made by the user. This could effectively aid localisation of the FRB, not spacially, but via distance.

\subsection{On the potential for disentanglement of FRB {\rm DM} components}

Figure \ref{FigHostRedshiftComp} shows DM probability distributions for FRBs arising from each of our progenitor scenarios. Results indicate distinguishable differences in potential DMs for FRBs arising from different models, particularly at low redshifts. This is most notable when comparing FRBs arising from stellar populations to those occurring close to galactic centers. This is a plausible result given the high electron densities and necessarily large ${\rm DM_{host}}$ contributions acquired from the latter. Indeed, in the most extreme cases where FRBs originate close ($<0.2$\,kpc) to galactic centers (see, e.g., \citet{urry95a, elit06a} for discussions on AGN structure) our models suggest this could become apparent for large numbers of FRBs binned by redshift, even out past $z>2$. When sufficient numbers of low-redshift FRBs have been detected, their DM distributions could be used to constrain progenitor scenarios. Conversely, as indicated by Fig. \ref{FigDMcuts}, techniques discussed here could be used to determine at what redshift ${\rm DM_{host}}$ is sufficiently suppressed as to be negligible for different progenitor scenarios.

Figure \ref{FigWMBComp} shows the probability of an FRB with the same DM as FRB121102 having a given redshift according various progenitors in spiral and elliptical galaxies, considering both homogeneous ($P(z_{\rm s})=1$) and Erlang distribution priors for FRB redshifts. This figure illustrates the noticeable affect of the host on redshift, and also that the underlying redshift distribution could noticeably affect the observed redshift distribution of a large number of FRBs when binned in DM.

With significant populations of FRBs at different redshifts, it is possible that individual DM components could be disentangled. It is expected that despite the true underlying $P({\rm DM_{exc}})$ being the same for FRBs at a certain source redshift, the total DM probability will display strong correlations as a function of sky direction due to the MW contribution. This will hold true even if the model MW contribution is subtracted as uncertainties with respect to the true value will remain. To mitigate this, future FRB surveys could consider two extreme approaches:
\begin{enumerate}
  \item Observations over a small sky area would reduce random ${\rm DM_{MW}}$ variations in the resulting FRB DM probability distribution. Over such an area the true MW contribution should be roughly uniform and would therefore result in a constant systematic uncertainty for all FRBs. Assuming FRBs have no preferential direction, this approach would yield a similar number of events as random pointing directions. FRB searches piggy-backing onto deep extra-galactic surveys would naturally fall within this class.

  \item Observations over a large sky area with a sufficiently large number of FRBs would enable to disentangle the ${\rm DM_{exc}}$ from the ${\rm DM_{MW}}$ contribution. Indeed, under the assumption that the probability of the former is direction-independent, the distribution of DMs for FRBs at a given redshift would display a direction-dependent contribution due to the MW. In fact, it will be possible to accomplish the same without redshift information in the case of a flux-limited survey as the overall DM PDF in any particular direction could be compared to another. Therefore, a survey like that planned for CHIME \citep{CHIME18} should provide enough FRBs to construct a sky map of the ${\rm DM_{MW}}$ contribution or implementing corrections to existing models. This is certainly an important area where detecting a large number of FRBs -- even with no/poor localisation -- is relevant, as the local DM contribution will have to be removed accurately in order to enable cosmological studies of the smaller sample of localised FRBs with known redshifts.
\end{enumerate}

Further, it will also be possible, in theory, to disentangle ${\rm DM_{IGM}}$ from ${\rm DM_{host}}$ given that the former is a function integrated over redshift whereas the latter is evaluated at the specific source redshift. Separating the two components could use techniques borrowed from cosmology such as those used to tell apart Galactic foreground emission from the CMB signal in Planck observations \citep{plan15a}. The main difference, however, is that even though the redshift dependence is known -- which in the CMB analogy would be the observed frequency -- the shape of the actual underlying PDFs will not be readily extractable. Component separation will likely require forward modelling using template IGM and host PDFs inspired from realistic models as presented in this paper, or a decomposition into basis functions such as a Gaussian mixture. Accomplishing this will necessitate redshifts to be available.

\subsection{Concluding remarks}

We present a framework for exploration of the statistical relationship between FRB redshifts and dispersion measures, which provides the basis for:

\begin{enumerate}
\item Qualitative assessment of host galaxy contributions to FRB DMs using realistic models. We find that all our host models may contribute large amounts of DM ($>400\;{\rm pc\;cm^{-3}}$) in the rest frame, and as expected, that ${\rm DM_{host}}$ is most significant for FRBs of lower source redshifts, becoming negligible as redshift increases. For the most extreme scenarios where FRBs originate close to galactic centers, this component still contributes significantly to overall $P(\rm{DM}|z_{\rm s})$ profiles out to $z_{\rm s}=3$.
\item More rigorous uncertainties to be placed on FRB redshifts than are currently standard practice. By consulting $P(z_{\rm s}|{\rm DM})$ probability distributions created from our (or similar) models, this may additionally provide an innovative way to narrow down the potential host galaxies for unlocalised FRBs, and allow insight into FRB progenitors to be drawn from large source populations. A repository containing our Python code and examples may be found online at \url{https://doi.org/10.5281/zenodo.1209920}.
\item The disentanglement of individual FRB dispersion measure components. For example, the MW components for given sightlines could be extracted from ${\rm DM_{obs}}$ by comparing DM probability distributions from a flux-limited survey (e.g. CHIME) at different sky locations and looking for  systematic offsets in their profiles. This technique would not require redshift measurements, thus further increasing the usefulness of unlocalised FRBs. It also could be possible to separate ${\rm DM_{IGM}}$ and ${\rm DM_{host}}$ using their respective redshift dependences.
\end{enumerate}

As more FRBs are localised and the research field continues to mature, the most likely host candidates for FRBs will become apparent, and accurate progenitor models may be employed accordingly.

\begin{acknowledgements}
      We thank Francesco Pace for vital conversation and instruction during this project, and Matthew McQuinn for providing useful comparison data for his IGM models. CRHW acknowledges support from a UK Science and Technology Facilities Council studentship. RPB acknowledges support from the ERC under the European Union's Horizon 2020 research and innovation programme (grant agreement no. 715051; Spiders).
\end{acknowledgements}


\bibliographystyle{aa} 
\bibliography{references} 

\begin{appendix}
\label{Ap1}
\section{Dispersion measurement and its variance from intergalactic medium} \label{sec:dispersion}
Section~\ref{sect:2.1}, Eq.\ref{eq:DMz}, provides the formula for the ${\rm DM_{IGM}}$, given FRB source redshift $z_{\rm s}$ \citep{mcqu14a}. Following Eqs. 3, 4, and Appendix (A1)--(A7) in~\citet{Ma14}, electron density can be written as;
\begin{eqnarray}
n_{\rm e}(z)=\frac{\chi_{\rm e} \rho_{\rm gas}(z)}{\mu_{\rm e}m_{\rm p}}, \label{eq:n-e}
\end{eqnarray}
where $\Omega_{\rm b} = 0.048$ is the fraction of baryon density~\citep{Planck_parameters}, $m_{\rm p}$ is the proton mass, $\mu_{\rm e}\simeq 1.14$ is the mean mass per electron, $\rho_{\rm gas}(z)=\rho_{\rm gas}(z=0)(1+z)^{3}=\Omega_{\rm b}\rho_{\rm cr}(1+z)^{3}$ is the gas density at redshift $z$, where $\rho_{\rm cr}=1.879\,h^{2}\times 10^{-29}\,{\rm g}\,{\rm cm}^{-3}$ is the critical density of the present Universe \citep{Dodelson}.  
\begin{equation}
\chi_{\rm e}=\frac{1-Y_{\rm p}(1-N_{\rm H_{e}}/4)}{1-Y_{\rm p}/2}
\end{equation}
Where $\chi_{\rm e}$ is the ratio between ionized and total number of electrons, $Y_{\rm p}\simeq 0.24$ is the helium mass fraction, and $N_{\rm H_{e}}$ is the number of ionized electrons per helium. The value of $N_{\rm H_{e}}$ can be between $0$ and $2$, so $\chi_{\rm e}$ ranges from $0.86$ to $1$.\\
Therefore we can write down our Eq. \ref{eq:DMz} more explicitly as:
\begin{eqnarray}
{\rm DM}(z_{\rm s}) = \left(\frac{\Omega_{\rm b}\chi_{\rm e}\rho_{\rm cr}}{m_{\rm p}\mu_{\rm e}} \right) \int^{z_{\rm s}}_{0}\der z \frac{(1+z)}{H(z)} .
\end{eqnarray}
Substituting the relevant quantities, we have:
\begin{eqnarray}
{\rm DM}(z_{\rm s})&=& 950.05\, [{\rm pc}\,{\rm cm}^{-3}]
\left(\frac{\Omega_{\rm b}}{0.048} \right)\left(\frac{\chi_{\rm e}}{1.0} \right)\left(\frac{h}{0.67} \right) \nonumber \\
& \times &\int^{z_{\rm s}}_{0}\der z \frac{(1+z)}{E(z)},
\end{eqnarray}
where $E(z)=\left[\Omega_{\rm m}(1+z)^{3}+\Omega_{\Lambda} \right]^{1/2}$.\\
Likewise, the variance of the sightline to sightline is~\citep{mcqu14a}:
\begin{eqnarray}
\sigma^{2}({\rm DM}) &=& \int^{\chi_{\rm s}}_{0}\der \chi (1+z)^{2}\bar{n}^{2}_{\rm e}(0) \int\frac{\der^{2}k_{\perp}}{(2\pi)^{2}}P_{\rm e}(k_{\perp},z) \nonumber \\
&=& \frac{c}{H_{0}}\left[\frac{\chi_{\rm e}\Omega_{\rm b}\rho_{\rm cr}}{\mu_{\rm e}m_{\rm p}} \right]^{2}\int^{z_{\rm s}}_{0}\der z\frac{(1+z)^{2}}{E(z)}\int \der k\, k P_{\rm e}(k,z), \nonumber \\
\end{eqnarray}
where:
\begin{eqnarray}
\bar{n}_{\rm e}(0)=\frac{\chi_{\rm e}\Omega_{\rm b}\rho_{\rm cr}}{\mu_{\rm e}m_{\rm p}},
\end{eqnarray}
as indicated by Eq.~(\ref{eq:n-e}).

\label{Ap2}
\section{Redshift constraints for existing FRBs}
\begin{table*}
\caption{DM-derived redshifts for catalogued FRBs (calculated via standard methods: $z\sim {\rm DM_{exc}}/(1200\,{\rm pc\,cm^{-3}})$ \citep{petr16a,ioka03a}), compared to redshifts derived using IGM-only and several of our host progenitor models (see Fig. \ref{FigHost}) with bounds drawn from their 99\% confidence intervals.}             
\label{FRBRedshiftTable}      
\centering          
\begin{tabular}{l l l l l l l }     
\hline\hline       
FRB & $z$ (DM-derived) & $z$ (IGM only) & $z$ (YPSR) & $z$ (H=0.2\,kpc) & $z$ $(R_{e}$=3.2\,kpc) & $z$ (H=0.1\,kpc)\\ 
\hline                    
      FRB170827 & 0.116 & $0.133^{+0.124}_{-0.130}$ & $0.088^{+0.123}_{-0.085}$ & $0.318^{+0.141}_{-0.315}$ & $0.083^{+0.124}_{-0.080}$ & $0.318^{+0.141}_{-0.315}$\\
   FRB150807 & 0.191 & $0.217^{+0.145}_{-0.215}$ & $0.173^{+0.142}_{-0.170}$ & $0.003^{+0.141}_{-0.000}$ & $0.152^{+0.220}_{-0.140}$ & $0.003^{+0.141}_{-0.000}$\\
   FRB160410 & 0.184 & $0.212^{+0.144}_{-0.210}$ & $0.173^{+0.170}_{-0.150}$ & $0.003^{+0.138}_{-0.000}$ & $0.152^{+1.180}_{-0.136}$ & $0.003^{+0.138}_{-0.000}$\\
   FRB010724 & 0.275 & $0.318^{+0.161}_{-0.315}$ & $0.278^{+0.271}_{-0.200}$ & $0.003^{+0.228}_{-0.000}$ & $0.253^{+1.235}_{-0.226}$ & $0.003^{+0.228}_{-0.000}$\\
   FRB130628 & 0.348 & $0.398^{+0.172}_{-0.395}$ & $0.357^{+1.260}_{-0.233}$ & $0.068^{+0.246}_{-0.065}$ & $0.318^{+1.285}_{-0.286}$ & $0.068^{+0.246}_{-0.065}$\\
   FRB120127 & 0.435 & $0.497^{+0.183}_{-0.495}$ & $0.458^{+1.310}_{-0.267}$ & $0.182^{+0.236}_{-0.180}$ & $0.422^{+1.320}_{-0.386}$ & $0.182^{+0.236}_{-0.180}$\\
   FRB121102 & 0.307 & $0.353^{+0.164}_{-0.350}$ & $0.307^{+0.224}_{-0.232}$ & $0.003^{+0.262}_{-0.000}$ & $0.288^{+1.245}_{-0.259}$ & $0.003^{+0.262}_{-0.000}$\\
   FRB140514 & 0.440 & $0.502^{+0.183}_{-0.500}$ & $0.458^{+0.272}_{-0.288}$ & $0.188^{+0.256}_{-0.184}$ & $0.422^{+1.305}_{-0.385}$ & $0.188^{+0.256}_{-0.184}$\\
   FRB170107 & 0.479 & $0.542^{+0.190}_{-0.540}$ & $0.502^{+0.464}_{-0.269}$ & $0.238^{+1.450}_{-0.215}$ & $0.468^{+1.320}_{-0.426}$ & $0.238^{+1.450}_{-0.215}$\\
   FRB110523 & 0.483 & $0.547^{+0.190}_{-0.545}$ & $0.507^{+1.320}_{-0.269}$ & $0.238^{+1.460}_{-0.214}$ & $0.468^{+1.330}_{-0.425}$ & $0.238^{+1.460}_{-0.214}$\\
   FRB160608 & 0.370 & $0.422^{+0.174}_{-0.420}$ & $0.378^{+0.246}_{-0.266}$ & $0.107^{+0.231}_{-0.105}$ & $0.343^{+1.290}_{-0.311}$ & $0.107^{+0.231}_{-0.105}$\\
   FRB110626 & 0.563 & $0.642^{+0.198}_{-0.640}$ & $0.597^{+0.273}_{-0.311}$ & $0.333^{+1.500}_{-0.300}$ & $0.537^{+1.410}_{-0.493}$ & $0.333^{+1.500}_{-0.300}$\\
   FRB010621 & 0.185 & $0.212^{+0.144}_{-0.210}$ & $0.173^{+0.170}_{-0.150}$ & $0.003^{+0.138}_{-0.000}$ & $0.152^{+1.180}_{-0.136}$ & $0.003^{+0.138}_{-0.000}$\\
   FRB150418 & 0.490 & $0.557^{+0.190}_{-0.555}$ & $0.512^{+0.259}_{-0.309}$ & $0.253^{+1.460}_{-0.229}$ & $0.473^{+1.345}_{-0.430}$ & $0.253^{+1.460}_{-0.229}$\\
   FRB131104 & 0.590 & $0.667^{+0.204}_{-0.665}$ & $0.627^{+0.300}_{-0.292}$ & $0.367^{+1.505}_{-0.332}$ & $0.573^{+1.395}_{-0.527}$ & $0.367^{+1.505}_{-0.332}$\\
   FRB010125 & 0.567 & $0.642^{+0.198}_{-0.640}$ & $0.597^{+0.273}_{-0.311}$ & $0.333^{+1.500}_{-0.300}$ & $0.537^{+1.410}_{-0.493}$ & $0.333^{+1.500}_{-0.300}$\\
   FRB130729 & 0.692 & $0.782^{+0.213}_{-0.780}$ & $0.737^{+0.271}_{-0.328}$ & $0.487^{+1.555}_{-0.439}$ & $0.672^{+1.475}_{-0.622}$ & $0.487^{+1.555}_{-0.439}$\\
   FRB090625 & 0.723 & $0.818^{+0.219}_{-0.815}$ & $0.777^{+0.295}_{-0.300}$ & $0.517^{+1.580}_{-0.465}$ & $0.717^{+1.485}_{-0.665}$ & $0.517^{+1.580}_{-0.465}$\\
   FRB110220 & 0.758 & $0.852^{+0.225}_{-0.850}$ & $0.812^{+0.273}_{-0.336}$ & $0.562^{+1.590}_{-0.505}$ & $0.742^{+1.510}_{-0.690}$ & $0.562^{+1.590}_{-0.505}$\\
   FRB130626 & 0.738 & $0.837^{+0.219}_{-0.835}$ & $0.792^{+0.264}_{-0.354}$ & $0.537^{+1.585}_{-0.482}$ & $0.717^{+1.510}_{-0.665}$ & $0.537^{+1.585}_{-0.482}$\\
   FRB151230 & 0.769 & $0.867^{+0.226}_{-0.865}$ & $0.828^{+0.280}_{-0.324}$ & $0.578^{+1.595}_{-0.518}$ & $0.747^{+1.525}_{-0.694}$ & $0.578^{+1.595}_{-0.518}$\\
   FRB110703 & 0.893 & $1.008^{+0.241}_{-1.005}$ & $0.967^{+0.275}_{-0.365}$ & $0.717^{+1.665}_{-0.576}$ & $0.867^{+1.610}_{-0.798}$ & $0.717^{+1.665}_{-0.576}$\\
   FRB150215 & 0.565 & $0.642^{+0.198}_{-0.640}$ & $0.597^{+0.273}_{-0.311}$ & $0.333^{+1.500}_{-0.300}$ & $0.537^{+1.410}_{-0.493}$ & $0.333^{+1.500}_{-0.300}$\\
   FRB170922 & 0.888 & $1.002^{+0.241}_{-1.000}$ & $0.962^{+0.274}_{-0.371}$ & $0.717^{+1.660}_{-0.582}$ & $0.867^{+1.595}_{-0.803}$ & $0.717^{+1.660}_{-0.582}$\\
   FRB160317 & 0.705 & $0.797^{+0.218}_{-0.795}$ & $0.762^{+0.388}_{-0.266}$ & $0.507^{+1.560}_{-0.457}$ & $0.672^{+1.490}_{-0.621}$ & $0.507^{+1.560}_{-0.457}$\\
   FRB171209 & 1.204 & $1.363^{+0.276}_{-1.360}$ & $1.323^{+0.289}_{-0.572}$ & $1.093^{+1.750}_{-0.643}$ & $1.162^{+1.730}_{-0.681}$ & $1.093^{+1.750}_{-0.643}$\\
   FRB150610 & 1.227 & $1.393^{+0.279}_{-1.390}$ & $1.357^{+0.315}_{-0.359}$ & $1.133^{+1.745}_{-0.662}$ & $1.162^{+1.765}_{-0.650}$ & $1.133^{+1.745}_{-0.662}$\\
   FRB121002 & 1.296 & $1.472^{+0.285}_{-1.470}$ & $1.432^{+0.301}_{-0.513}$ & $1.212^{+1.760}_{-0.684}$ & $1.242^{+1.785}_{-0.652}$ & $1.212^{+1.760}_{-0.684}$\\
   FRB151206 & 1.458 & $1.667^{+0.304}_{-1.665}$ & $1.633^{+0.329}_{-0.418}$ & $1.412^{+1.805}_{-0.727}$ & $1.442^{+1.840}_{-0.642}$ & $1.412^{+1.805}_{-0.727}$\\
   FRB160102 & 2.153 & $2.578^{+0.417}_{-0.384}$ & $2.553^{+0.543}_{-0.365}$ & $2.353^{+1.965}_{-0.614}$ & $2.373^{+2.005}_{-0.590}$ & $2.353^{+1.965}_{-0.614}$\\   \hline                  
\end{tabular}
\end{table*}

\end{appendix}

\end{document}